\documentclass[floatfix,preprintnumbers,amsmath,amssymb,aps,twocolumn,prl,10pt]{revtex4}

\usepackage{amsfonts}
\usepackage{amsmath}
\usepackage[pdftex]{graphicx}
\usepackage[pdftex,usenames]{color}
\usepackage{bm}
\usepackage{multirow}

\begin{document}

\title{Adiabatic preparation of vortex lattices}
\author{Stefan K. Baur}
\affiliation{T.C.M. Group, Cavendish Laboratory, J. J. Thomson Avenue, Cambridge CB3 0HE, United Kingdom}
\author{Nigel R. Cooper}
\affiliation{T.C.M. Group, Cavendish Laboratory, J. J. Thomson Avenue, Cambridge CB3 0HE, United Kingdom}

\date{\today}
  
\begin{abstract}
  By engineering appropriate artificial gauge potentials, a
  Bose-Einstein condensate (BEC) can be \emph{adiabatically} loaded
  into a current carrying state that resembles a vortex lattice of a
  rotating uniform Bose gas. We give two explicit, experimentally
  feasible protocols by which vortex lattices can be smoothly
    formed from a condensate initially at rest. In the first example
  we show how this can be achieved by adiabatically loading a
  uniform BEC into an optical flux lattice, formed from coherent
    optical coupling of internal states of the atom. In the second
  example we study a tight binding model that is continuously
  manipulated in parameter space such that it smoothly transforms into
  the Harper-Hofstadter model with $1/3$ flux per plaquette.
\end{abstract}
\maketitle

\section{Introduction}

One of the most striking signatures of the quantum coherence of a
Bose-Einstein condensate (BEC) is the formation of a lattice of
quantized vortices when the condensate is forced to
rotate~\cite{fetter,advances}.  In the rotating frame of reference, the
particles experience a Coriolis force which plays the same role as the
Lorentz force on a charged particle in a uniform magnetic
field. Hence, the vortex lattice of a rotating BEC has similar origin
to the vortex lattice of a superconductor in a magnetic field.  The
density of vortex lines is set by the (effective) magnetic flux
density, which is $n_\phi = 2M \Omega/h$ for atoms of mass $M$
rotating at angular frequency $\Omega$.

There are two common ways in which to form a vortex lattice.  In one
method, commonly used in superconductors, one starts from the normal
(uncondensed) phase already subjected to the magnetic field. On
cooling, the system undergoes a phase transition directly into the
superfluid (condensed) phase with a vortex lattice of non-zero density
$n_\phi\neq 0$.  In another method, commonly used for liquid
helium~\cite{donnelly} and for dilute atomic gases, the system is first
cooled into the condensed phase in the absence of any effective
magnetic field, $n_\phi=0$.  The effective magnetic field is then
gradually increased, for example by imposing a rotating deformation.
As the field strength increases, vortices must enter from outside the
condensate~\cite{Madison:2000kx,Raman:2001vn,Lin:2009qf,
  Sinha:2001uq}. For atomic BECs, this is achieved via surface wave
instabilities
 and involves an interesting and complex dynamical
evolution~\cite{dalfovostringari,PhysRevLett.86.564, PhysRevLett.92.020403}, including periods in which the vortex lattice is highly
disordered and far from equilibrium.  Nevertheless, by transfer of
energy from the disordered vortex lattice into phonon modes
(i.e. heating of the BEC) or by additional evaporative cooling, the
system can be stabilised into ordered arrays of vortices.  This has
been shown in various experiments, using rotation or Raman coupling to
generate the effective magnetic flux density~\cite{Madison:2000kx,Raman:2001vn,SchweikhardCEMC92,Lin:2009qf}.

In this paper we describe a novel alternative route to creating a
dense vortex lattice in an atomic BEC: by adiabatic manipulation of
optical lattice potentials with artificial gauge fields.  Recently,
new ways to create strong magnetic
fields for cold atoms have been put forward, and are now within reach
of experiments~\cite{Jaksch:2003vn,Gerbier:2010kl,Cooper:2011kl,
  Cooper:2011cr,Cooper:2013ve,Juzeliunas,
  Kolovsky:2011zr, Aidelsburger:2011kx, Aidelsburger:2012,Struck,Jimenez,
  Hauke:2012fk}.  These methods lead to very high flux densities,
about two orders of magnitude larger than previous experimental works.
Thus, following the standard approach for cold gases and increasing
the flux density from $n_\phi =0$ would require a very large number of
vortices to enter the system, potentially driving the system very far
from equilibrium and requiring significant cooling to maintain
BEC. Our proposal shows that these very dense vortex arrays can in
fact be formed {\it adiabatically}, maintaining the system at
ultracold temperatures without requiring any further cooling.

We describe two generic experimental protocols by which a vortex
lattice can be adiabatically created from a uniform BEC. The first
setup involves loading a BEC into an optical flux lattice\cite{Cooper:2011kl,Cooper:2011cr,
Juzeliunas,Cooper:2013ve}, based on
the coherent (Rabi) coupling of internal atomic states. We describe
the density and current patterns in the system following loading, and
show that these are as expected for the dense vortex lattice.
In the second part of the paper, we turn to consider the formation of
vortex lattices in the Harper-Hofstadter model\cite{harper,hofstadter} for atoms moving on a
tight-binding lattice.  We present an experimental protocol by which
the uniform BEC for vanishing flux per plaquette can be adiabatically
transformed into the vortex-lattice ground-state of a lattice with
$1/3$ flux per plaquette.
\section{Optical flux lattice}

We consider bosonic atoms with two internal states, which we label by
the (pseudo)-spin $\uparrow$ and $\downarrow$. The atoms are subjected
to coherent optical fields which, in the rotating wave approximation,
are described by the potential $\hat{V}(\mathbf{r}) = \sum_{i=x,y,z}
A_i(\mathbf{r})\hat{\sigma}_i$ with $\hat{\sigma}_{x,y,z}$ the Pauli
matrices acting on the internal states. The amplitudes $A_i(\mathbf{r})$
describe the strengths of the local optical coupling of the two
internal levels ($A_{x,y}$) and of a state-dependent potential
($A_z$).  Various implementations of such couplings are possible,
using electronic states, hyperfine levels or even vibrational
states~\cite{dalibardreview,Parker:2013zr}.  When the optical coupling is dominant
(compared to the kinetic energy $E_{\rm L}$ defined below), the internal state of
the atom is restricted to the lowest energy 
eigenstate of $\hat{V}$,
which we denote by the dressed state $|0_\mathbf{r}\rangle =
\alpha_\mathbf{r}|\uparrow\rangle+\beta_\mathbf{r}|\downarrow\rangle$. In
this limit, the atom moves through space adiabatically with overall
wavefunction $|\Psi(\mathbf{r})\rangle = \psi_0(\mathbf{r}) |0_{\mathbf{
 r}}\rangle$. The Berry curvature associated with spatial variations
of the dressed state $| 0_\mathbf{r}\rangle$ causes the motion of the atom, as
described by the positional wavefunction $\psi_0(\mathbf{r})$, to
experience an effective magnetic field~\cite{dalibardreview}.

An optical flux lattice is a periodic configuration of the optical
fields which cause the atom to experience a non-zero number of flux
quanta, $N_\phi\neq 0$, per unit cell~\cite{Cooper:2011kl}.  We focus
on a simple, but representative, example of an optical flux lattice,
introduced in Ref.~\onlinecite{Cooper:2011kl}
\begin{equation}
\label{eq:ofl}
\hat{V}(x,y)=V_0 \left[ \hat{\sigma}_x \cos(\mathbf{k}_1 \cdot \mathbf{r})+ \hat{\sigma}_y \cos(\mathbf{k}_2 \cdot \mathbf{r}) +\hat{\sigma}_z \cos(\mathbf{k}_3 \cdot \mathbf{r}) \right],
\end{equation}
with $\mathbf{k}_1=\frac{2 \pi}{a} (2/\sqrt{3},0)$,
$\mathbf{k}_2=\frac{2 \pi}{a} (1/\sqrt{3},1)$, and
$\mathbf{k}_3=\mathbf{k}_2-\mathbf{k}_1$ defining three reciprocal lattice vectors 
in the $xy$ plane.  In real space, this lattice has triangular symmetry, with lattice vectors $\mathbf{a}_1 = a(0,1)$ and $\mathbf{a}_2 =
a(\sqrt{3}/2,1/2)$. Within this unit
cell, the lowest energy dressed state experiences $N_\phi=2$ flux
quanta.  Thus, one expects that a BEC in this lowest energy dressed
state will exhibit $N_\phi=2$ vortices per unit cell.

\begin{figure}
\includegraphics[width=0.9\columnwidth]{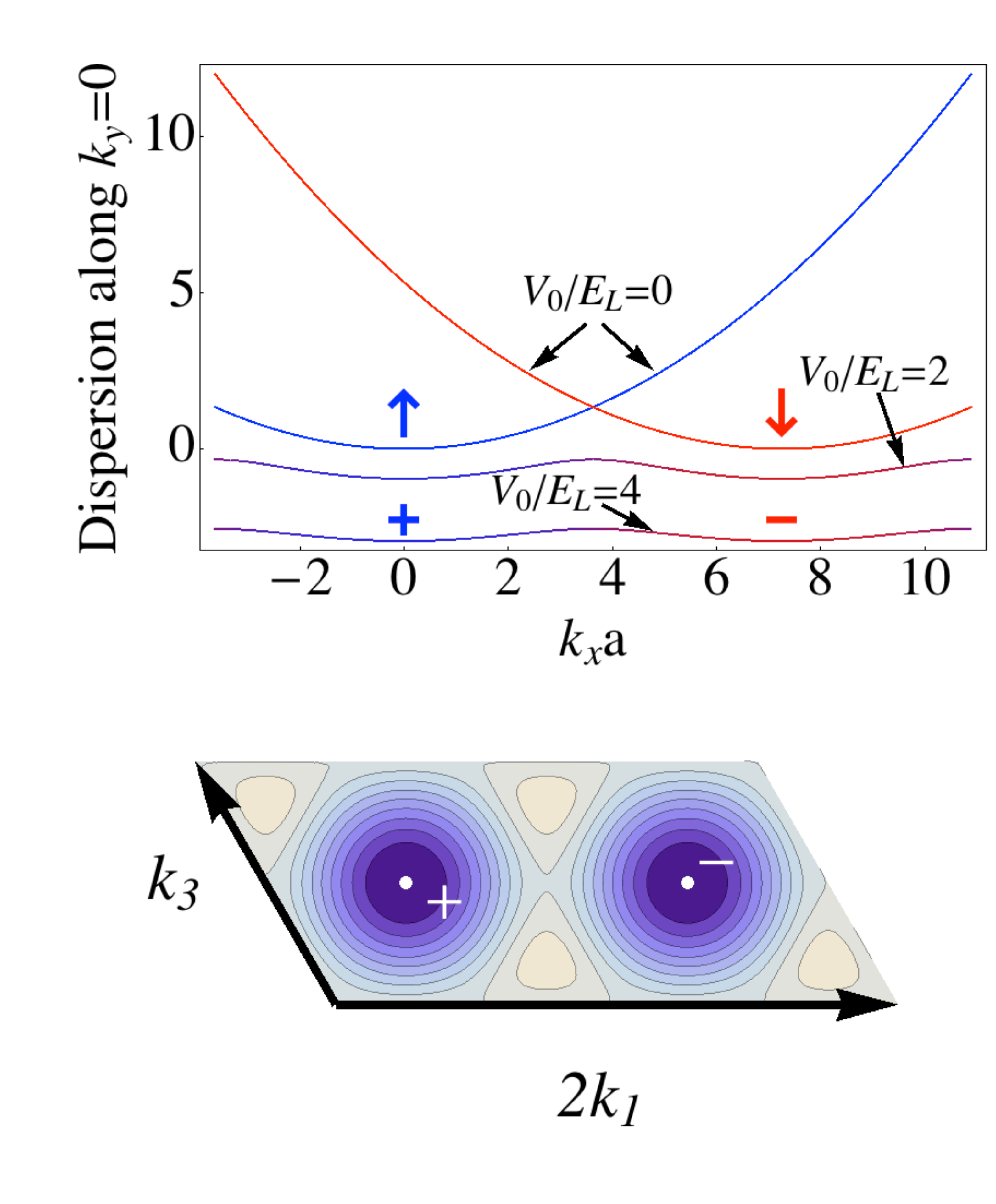}
\caption{(Color online) Cut through the dispersion of the lowest band in the optical flux lattice along $k_x$ with $k_y=0$ (passing through the minima) for lattice depths $V_0/E_{\rm L}=0,2,4$ (top). The dressed states of the lowest band are superpositions of spin-$\uparrow$ (shown in blue) and spin-$\downarrow$ (red) states. The dispersion y-axis has units of $E_{\rm L}$.  The dispersion (bottom) shown here is for a lattice depth of $V_0/E_{\rm L}=4$.
}
\label{fig:condensate}
\end{figure}

The energy bands follow from the eigenstates of the Hamiltonian including the
kinetic energy, 
$\hat{H} = \frac{\mathbf{p}^2}{2M}\hat{\openone}_2 + \hat{V}(\mathbf{r})$.
(We focus on the motion in the $xy$ plane; motion normal to this plane, along $z$, remains
free particle-like and in the Bose-condensed phases we describe the
atoms will simply condense in $p_z=0$ state.)  
The 
bandstructure depends on the lattice depth $V_0/E_{\rm L}$ where $E_{\rm L}
\equiv \hbar^2 \pi^2/(2M a^2)$. A cut through the energy bands is shown in
Fig.~\ref{fig:condensate} (top) for three values of $V_0/E_{\rm L}$. At all lattice depths,
the single particle states have two degenerate minima. This degeneracy
is a consequence of the discrete symmetry
operations~\cite{Cooper:2011kl}
\begin{eqnarray}
\hat{T}_1=\hat{\sigma}_z e^{(\mathbf{a}_1/2)\cdot \mathbf{\nabla}} \;\;,\;\; \hat{T}_2=\hat{\sigma}_x e^{(\mathbf{a}_2/2)\cdot\mathbf{\nabla}} 
\end{eqnarray}
involving translations by $\mathbf{a}_1/2$ and $\mathbf{a}_2/2$ combined
with spin rotations, for which $\hat{T}_1 \hat{T}_2=-\hat{T}_2
\hat{T}_1$. Since $[\hat{T}_1,\hat{T}_2^2]=0$, energy eigenstates can
be made simultaneous eigenfunctions of $\hat{T}_1$ and $\hat{T}_2^2$,
so the magnetic unit cell can be chosen to have sides $\mathbf{a}_1/2$
and $\mathbf{a}_2$, containing $N_\phi=1$ flux quantum. This leads to the
magnetic Brillouin zone in Fig.~\ref{fig:condensate} (bottom), with
reciprocal lattice spanned by $2\mathbf{k}_1$ and $\mathbf{k}_3$.  
The states of minimum kinetic energy, which we label $\pm$, are
located at $\mathbf{k}_+= 0$ and $\mathbf{k}_-=\mathbf{k}_1$. 
These
continuously evolve into the eigenstates of $\hat{\sigma}_z$ with zero
kinetic energy as $V_0/E_{\rm L} \rightarrow 0$ ~\cite{footnote1}.

This continuous evolution of the bandstructure with varying
$V_0/E_{\rm L}$ allows the adiabatic preparation of a BEC in the
minima of the lowest band of the optical flux lattice.  We consider a
protocol where the lattice lasers are ramped up slowly from $V_0=0$,
thereby loading a weakly interacting BEC into the lattice. As the
Rabi coupling is mixing the spin degrees of the two-component BEC, for
adiabaticity, we have to ensure that as the lattice is turned on, the
BEC remains in the mean-field ground-state.
Consider, first,
non-interacting bosons in an ideal infinite (untrapped) system. Let us
start with a BEC of spin-$\uparrow$ atoms, that is with condensate
wave function
\begin{eqnarray}
\phi_{\rm i}=\sqrt{n_0} \left( \begin{array}{c} 1 \\ 0 \end{array} \right),
\label{eq:becup}
\end{eqnarray}
where $n_0=N/A$ is the number of atoms $N$ per area $A$. Now,
increasing $V_0/E_{\rm L}$ from zero will causes the condensate
wavefunction to evolve continuously into that of the $\mathbf{k}_+$
state, thereby adiabatically loading the atoms into a BEC in this 
minimum.  Similarly, a BEC in any initial superposition of
spin-$\uparrow$ and spin-$\downarrow$ will evolve into a BEC in a
superposition state of the degenerate minima at $\mathbf{k}_\pm$.
For a finite system in a trap, when $V_0$ is nonzero the trap
potential can cause scattering of particles between the two degenerate
minima.  Then, other considerations are required in order to ensure
adiabatic loading of the BEC.
One  way to achieve adiabaticity is to detune the laser(s)
providing the Rabi coupling from resonance by an amount
$\delta$. This adds a spatially uniform term $-(\hbar\delta/2)
\hat{\sigma}_z$ to the optical coupling (\ref{eq:ofl}) which breaks the
degeneracy of the two minima. The lowest energy band has a single
non-degenerate minimum for all lattice depths, which may be adiabatic
loaded without sensitivity to scattering processes.
Alternatively, one can make use of the fact that inter-particle
interactions can lift the degeneracy of BECs with two internal states.
Specifically, we consider the effects of state-dependent interactions,
for which the mean-field interaction energy is given by
\begin{eqnarray}
E_{\rm int}=\int d^2r\; \frac{g_{\uparrow \uparrow}}{2} n_{\uparrow}^2(\mathbf{r})+ \frac{g_{\downarrow \downarrow}}{2} n_{\downarrow}^2(\mathbf{r})+g_{\uparrow \downarrow} n_{\uparrow}(\mathbf{r}) n_{\downarrow}(\mathbf{r}),
\end{eqnarray}
where $n_{\uparrow,\downarrow}(\mathbf{r})$ are the
spin-$\uparrow/\downarrow$ densities of the condensate wave functions
and $g_{\uparrow \uparrow}$/$g_{\downarrow \downarrow}$ ($g_{\uparrow
  \downarrow}$) are the intra- (inter-) species interactions.  Under
the assumption of weak interactions, the condensate wave function is a
linear combination of the two degenerate minima, $\pm$.
The relative sizes of the state-dependent interactions determine the
spin-state of the lowest energy BEC.  For simplicity, consider the
regime where
\begin{eqnarray}
\label{eq:ints}
g_{\uparrow \downarrow}>g_{\downarrow \downarrow}>g_{\uparrow \uparrow}>0.
\end{eqnarray}
Then, for $V_0=0$ the lowest energy BEC involves a condensate with
only $|\uparrow\rangle$, as in Eq (\ref{eq:becup}).  This condensate
wave function minimizes the interaction energy of the free Bose
gas. As above, the condensed state continuously evolves with
increasing $V_0/E_{\rm L}$ remaining the mean-field ground-state of
the lattice potential.
\begin{figure}
\includegraphics[width=0.9 \columnwidth]{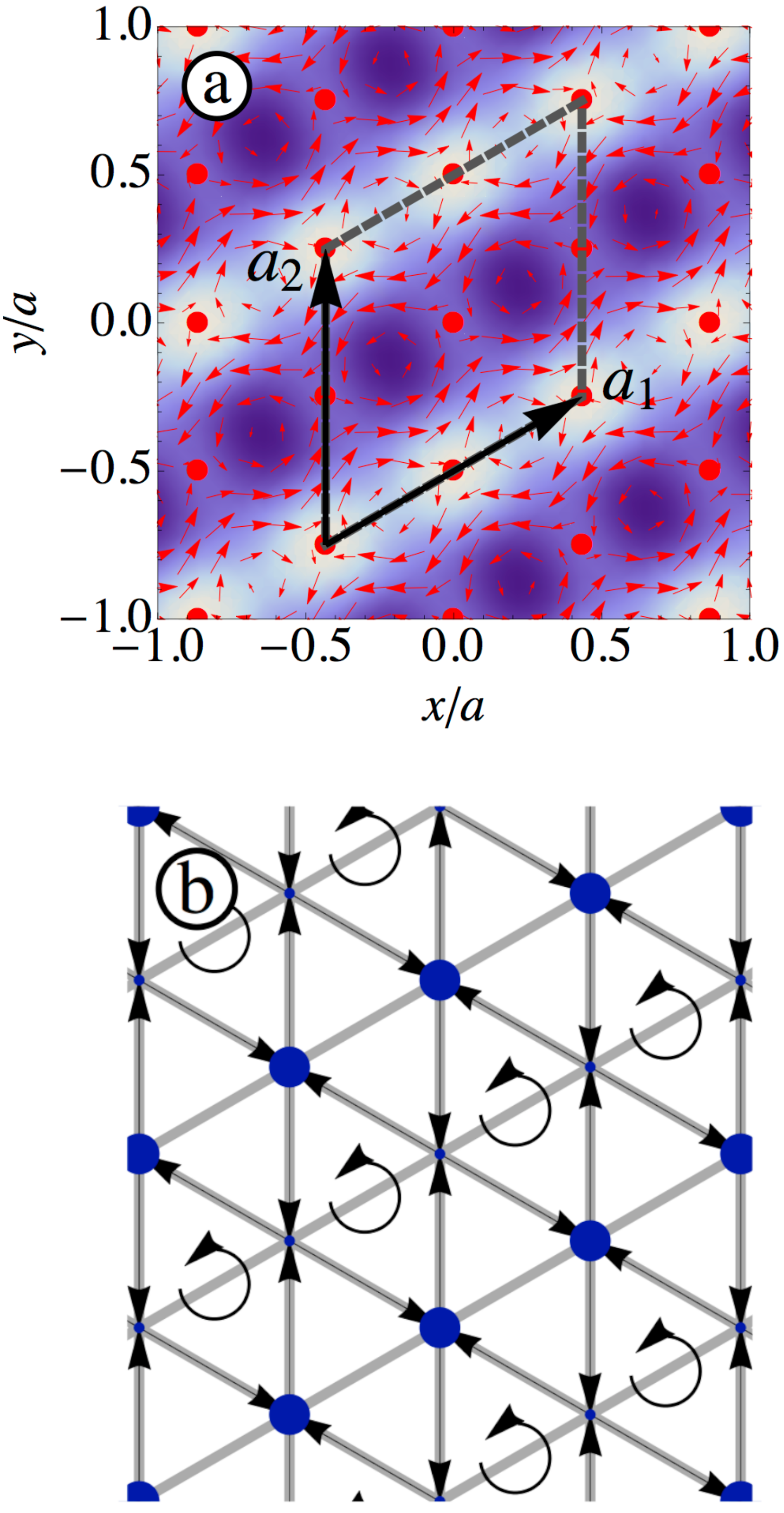}
\caption{(Color online) (a) Total density and current density 
for the condensate wavefunction described in the main text at lattice depth $V_0/E_{\rm L}=4$ [light (dark) colors correspond to high (low) density].  The arrows denote the Bravais lattice vectors $\mathbf{a}_1, \mathbf{a}_2$ of the flux lattice. In the deep lattice limit, the atoms are localized at the lattice sites marked by the red dots. (b) Density of the analog condensate wave function for the corresponding tight-binding lattice model (large blue dots indicate high density). The arrows indicate the direction of mass currents (all currents have same magnitude).}\label{fig:triangularflux}
\end{figure}

In Fig.~\ref{fig:triangularflux} we show the condensate wavefunction
formed by adiabatically loading a BEC into the $+$-minimum, to a
lattice depth of $V_0/E_{\rm L} = 4$.  This Figure shows both the
particle density (shading) and the current density (arrows). An
inspection of the pattern of densities and currents shows that these
have the expected features of a vortex lattice.  In the unit cell of
sides $\mathbf{a}_{1,2}$ the lowest energy dressed state experiences
$N_\phi=2$ flux quanta, so we expect that there should appear 2
quantized vortices. Indeed, clear signatures of these 2 vortices
appear: there are two points around which the current circulates (in
an anticlockwise sense) and at the centre of which the particle
density falls to a small value.  There are also two stagnation points,
around which the current density circulates in a clockwise
sense. These are required by periodicity of the flow field (in the
rest frame of the vortex lattice there is no net flow), so appear also
for a rotating superfluid. They are not quantized vortices (or
antivortices) since the particle density remains large at the centres
of these points, so the velocity field is regular and has zero net
circulation around these points.

It is clear from Fig.~\ref{fig:triangularflux} that the vortices do
not form a triangular lattice, familiar for rotating BECs.  Rather,
the vortices are arranged in a rectangular array.  This is due to the
fact that the dominant energy is the lattice potential, so the
vortices arrange in order to minimize the energy of the optical
coupling (\ref{eq:ofl}).  The rectangular arrangement of the vortices
leads to a particle density with a stripe-like variation in the
direction perpendicular to the vector $\mathbf{a}_1$.  This reflects
the fact that, when condensed in the Bloch wavefunction at the $+$
minimum, the atoms have large magnetization along the
$z$-direction. The energy of the optical coupling (\ref{eq:ofl}) is
minimized by the pinning of the density wave with density maxima along
lines where $\cos(\mathbf{k}_3 \cdot \mathbf{r})=-1$.

Since the formation of the vortex is adiabatic, as the lattice depth
is ramped up the density modulation and current pattern both grow
smoothly and continuously, starting from uniform density and vanishing
current for $V_0 =0$.
\begin{figure}
\includegraphics[width=0.8 \columnwidth]{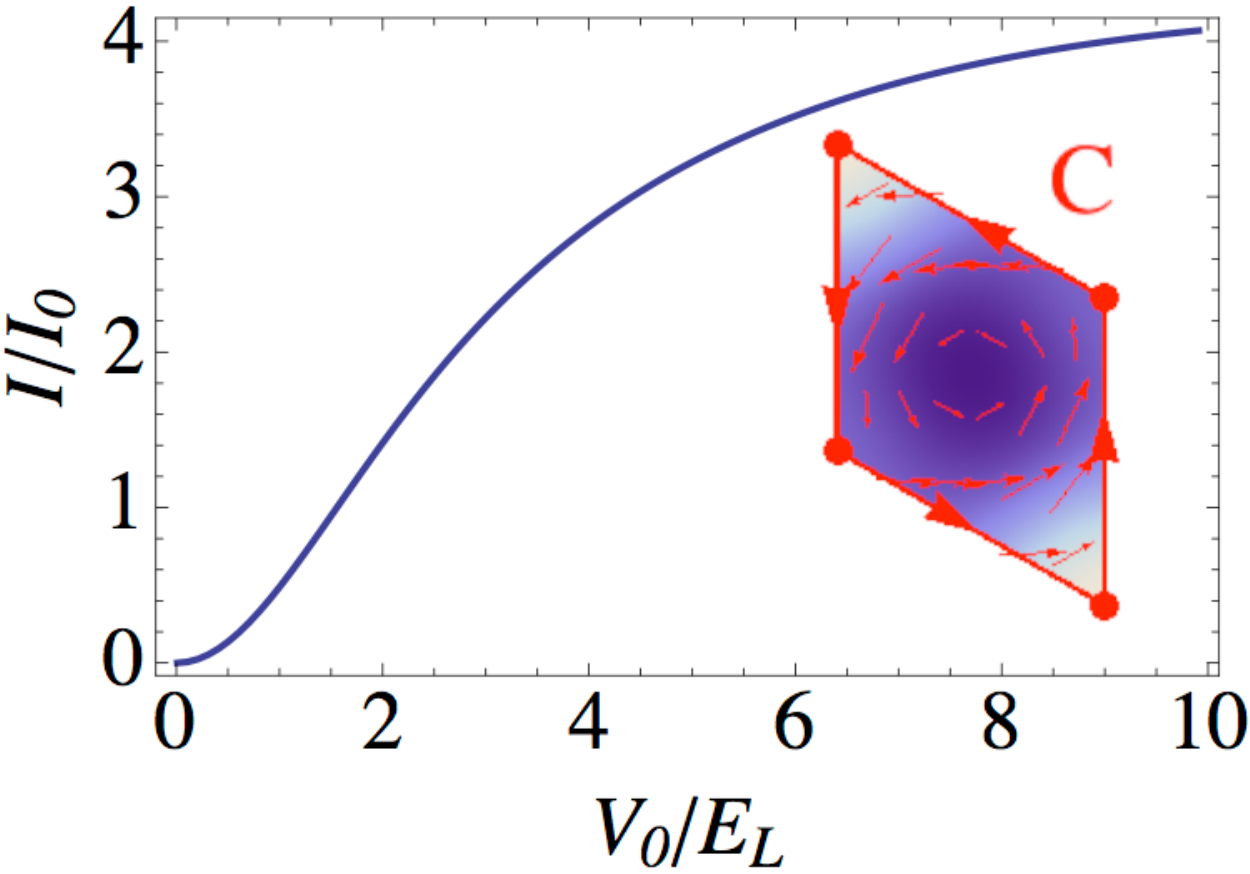}
\caption{(Color online) As the depth of this optical flux lattice is ramped up, currents appear smoothly in the condensate. We quantify the mass flow by calculating the line integral $I$ [Eq. \eqref{eq:lineintegral}] over total current density $\mathbf{j}=\mathbf{j}_{\uparrow}+\mathbf{j}_{\downarrow}$ along the contour shown in the inset. We have normalized $I$ with $I_0=\hbar n_0/M$.}\label{fig:currentsflux}
\end{figure}
To quantify the mass flow in the flux lattice, we study the (gauge invariant) total mass current density
\begin{eqnarray}
\mathbf{j}(\mathbf{r})=\mathbf{j}_{\uparrow}(\mathbf{r})+\mathbf{j}_{\downarrow}(\mathbf{r}),
\end{eqnarray}
where
$\mathbf{j}_{\uparrow/\downarrow}(\mathbf{r})=\hbar/M\; {\rm Im}\left[ \psi_{\uparrow/\downarrow}^{*}(\mathbf{r}) \partial_{\mathbf{r}} \psi_{\uparrow/\downarrow}(\mathbf{r})\right]$. To demonstrate that currents smoothly increase from zero, we plot as a measure of flow the line integral of $\mathbf{j}$ along the edges of the contour $C$ (see the inset to Fig. \ref{fig:currentsflux})
\begin{eqnarray}
\label{eq:lineintegral}
I= \oint_C d\mathbf{r}\cdot \mathbf{j}(\mathbf{r}).
\end{eqnarray}
As can be seen in Fig. \ref{fig:currentsflux}, the currents increase continuously from zero as the lattice depth is increased, thus demonstrating the adiabatic creation of a vortex lattice. 

How can a vortex lattice of fixed density ($N_\phi =2$ per unit cell)
build up continuously?  This might seem impossible.  After all, recall
that a quantized vortex is associated with a singularity at the vortex
core (point-like in two-dimensions or line-like in three dimensions)
at which the condensate density vanishes.  How can one smoothly
transform from a uniform BEC into a vortex lattice with zeroes in the
density?  The resolution lies in the fact that it is only for a {\it
  one}-component superfluid that the vortex core need have vanishing
density.  For a two- (or more-) component superfluid it is possible
for the particle density to remain non-zero everywhere, in so-called
``coreless vortices''~\cite{Mermin:1976ys, Ho:1998fk,Ohmi:1998kx}.
For the optical flux lattice, in general the condensate has a
two-component wavefunction, which we may write
$|\Psi(\mathbf{r})\rangle = \psi_0(\mathbf{r})|0_\mathbf{r}\rangle +
\psi_1(\mathbf{r})|1_\mathbf{r}\rangle$ in terms of the two dressed
states ($|0_\mathbf{r}\rangle$ and $|1_\mathbf{r}\rangle$). For any
finite lattice depth $V_0/E_{\rm L} < \infty$ this two-component
condensate has coreless vortices, so the density does not vanish at
the vortex core. (In Fig.~\ref{fig:currentsflux}, $V_0/E_{\rm L} = 4$
is finite so, although strongly suppressed, the density remains
non-zero at the vortex cores.)  As the lattice depth $V_0/E_{\rm L}$
is increased, the density suppression at the vortex core gradually
develops. In the limit $V_0/E_{\rm L}\gg 1$ for which $|\Psi(\mathbf{r})\rangle = \psi_0(\mathbf{r})|0_\mathbf{r}\rangle $, the
condensate is a one-component function $\psi_0(\mathbf{r})$ which must
have zeros at the vortex core.

In addition to the developing vortex core, in the limit $V_0/E_{\rm
  L}\gg 1$ the scalar potential experienced by the lowest energy
dressed state $|0_{{\mathbf{r}}}\rangle$ causes the atoms to become tightly
confined to lattice sites of a triangular lattice (with spacings $\mathbf{a}_{1,2}/2$). In this limit, the optical flux lattice maps onto a
triangular tight-binding lattice model with $1/4$ flux per
plaquette~\cite{Cooper:2011kl}. The tight-binding limit of the
condensate wavefunction in the optical flux lattice is shown in
Fig. \ref{fig:triangularflux} (b). The vortices reside along rows of
reduced density as marked by the arrows with one vortex per 4 lattice
sites.

\section{Tight binding model} 
We will now describe a complementary protocol for adiabatically transforming a condensate in a tight binding lattice into a vortex lattice.  

Now consider a condensate subjected to a deep optical lattice (without
any applied artificial gauge potentials), in such a way that the
atomic motion is well described by hopping between states localized at
the lattice sites. As we will show below, by turning on appropriate
photon assisted hoppings~\cite{Eckardt:2005tg} between nearest
neighbour lattice sites, the mean-field ground-state can be smoothly
evolved into the ground-state of the Harper-Hofstadter
model. We will focus on a square lattice with
$1/3$ flux per plaquette. As a consequence of magnetic translation
invariance, the single particle states for a lattice with $p/q$ flux
per plaquette are $q$-fold degenerate. The mean-field condensate wave
function for a weakly interacting BEC is a linear superposition of the
Bloch states at the $q\equiv 3$ minima of the dispersion relation. The
(infinite) degeneracy of all different superposition states is lifted
by interactions. We assume that interactions are sufficiently
weak that the atoms only occupy states in the three degenerate minima.
Minimizing a mean-field on-site repulsive interaction $E_{\rm
  int}=\frac{U}{2} \sum_i n_i (n_i-1)$ favours a condensate with
uniform density, giving rise to a ground-state with rows of vortices
along the diagonal of the square lattice with one vortex per three
lattice sites~\cite{Straley:1993fk,Powell:2010fk,Powell:2011fk}. The infinite degeneracy is lifted, and replaced by a
residual six-fold degeneracy, arising from transformations of the
vortex lattice configuration by translations and rotations by
$90^{\circ}$~\cite{Zhang:2010uq}.

\begin{figure}
\includegraphics[width=0.8\columnwidth]{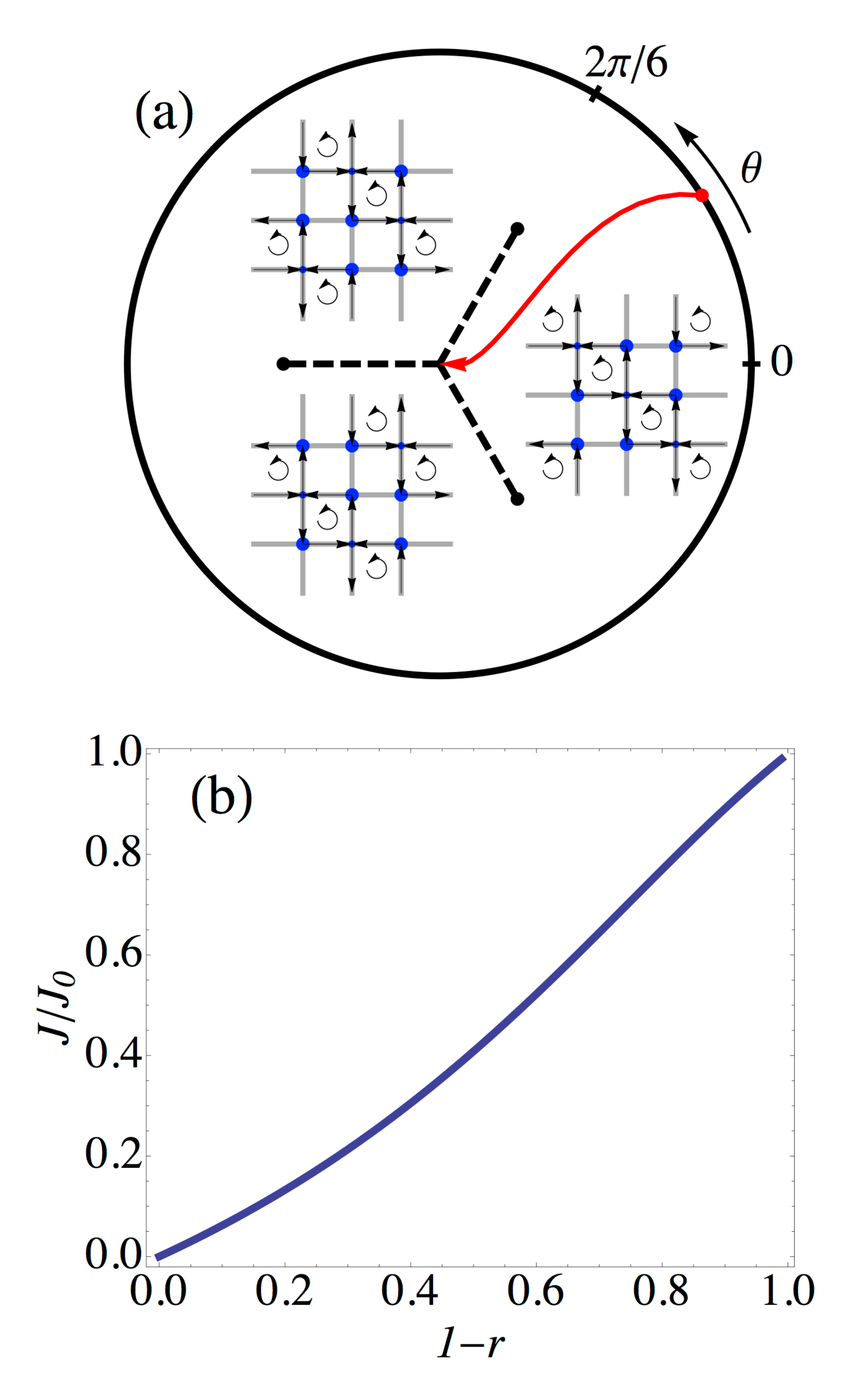}
\caption{(Color online) (a) When adiabatically ramping with the hopping matrix elements of Eq. \ref{eq:rampprotocol}, the angle $\theta$ selects between different translations of the vortex lattice. To be adiabatic, coming from $r=1$ (circle), one has to avoid crossing the dashed lines when approaching $r=0$ (centre) as shown for example by the solid red path. (b) Normalized average current for the square lattice as a function of $r$ along $\theta=0$.}\label{fig:currentstb}
\end{figure}

Our goal is to describe a protocol by which smooth variations of
experimentally controlled parameters adiabatically transform a
condensate in the lattice without gauge potential into the mean-field
ground states shown in Fig. \ref{fig:currentstb}
(a).
Care is required to ensure that the adiabatic route takes the system
directly into one of the six (degenerate) ground states that are
favoured by repulsive interaction. We achieve this by following a
route which breaks translational symmetry in such a way that the
system is guided directly into a chosen vortex lattice configuration.
To this end, we consider a square lattice tight binding model with Hamiltonian
\begin{equation}
H=-\sum_{n,m} K_{n,m} a_{n,m}^{\dagger} a_{n+1,m} -K\sum_{n,m} a_{n,m}^{\dagger} a_{n,m+1}+{\rm h.c.} ,
\end{equation}
where $a^{(\dag)}_{n,m}$ are bosonic destruction (creation) operators
with the integers $(n,m)$ labelling the site in the $(x,y)$ directions
respectively. For flux $1/3$ per plaquette, the hopping matrix elements along $x$ can be chosen as
 \begin{eqnarray}
 \label{eq:matrixelments}
K_{n,m}^{0}=\left\{ \begin{array}{c} K \;\;\; m+n=0,3,\ldots\\ K e^{-i 2\pi/3} \;\; m+n=1,4,\ldots \\ K e^{i 2\pi/3} \;\;\; m+n=2,5,\ldots \end{array} \right. ,
\end{eqnarray}
with real $K>0$ setting also the (uniform)  hopping matrix elements along the $y$-axis. This particular gauge is the most straightforward to implement experimentally when using photon assisted tunneling as described in Refs. \cite{Kolovsky:2011zr,Aidelsburger:2012}. 
Additional control of the tunneling matrix elements can be achieved by combining this with a second source of photon-assisted hopping, but with spatially uniform phase pattern~\cite{blochgroup}, which 
can be achieved by shaking the lattice along the $x$-axis~\cite{Sias:2008ly} or alternatively by lattice modulation. The combined effects lead to net tunneling matrix elements
\begin{eqnarray}
\label{eq:rampprotocol}
K_{n,m} = K e^{i \theta} r+(1-r)K_{n,m}^{0},
\end{eqnarray}
where $\theta$ is the relative phase of the two drives.

\begin{figure}
\includegraphics[width=\columnwidth]{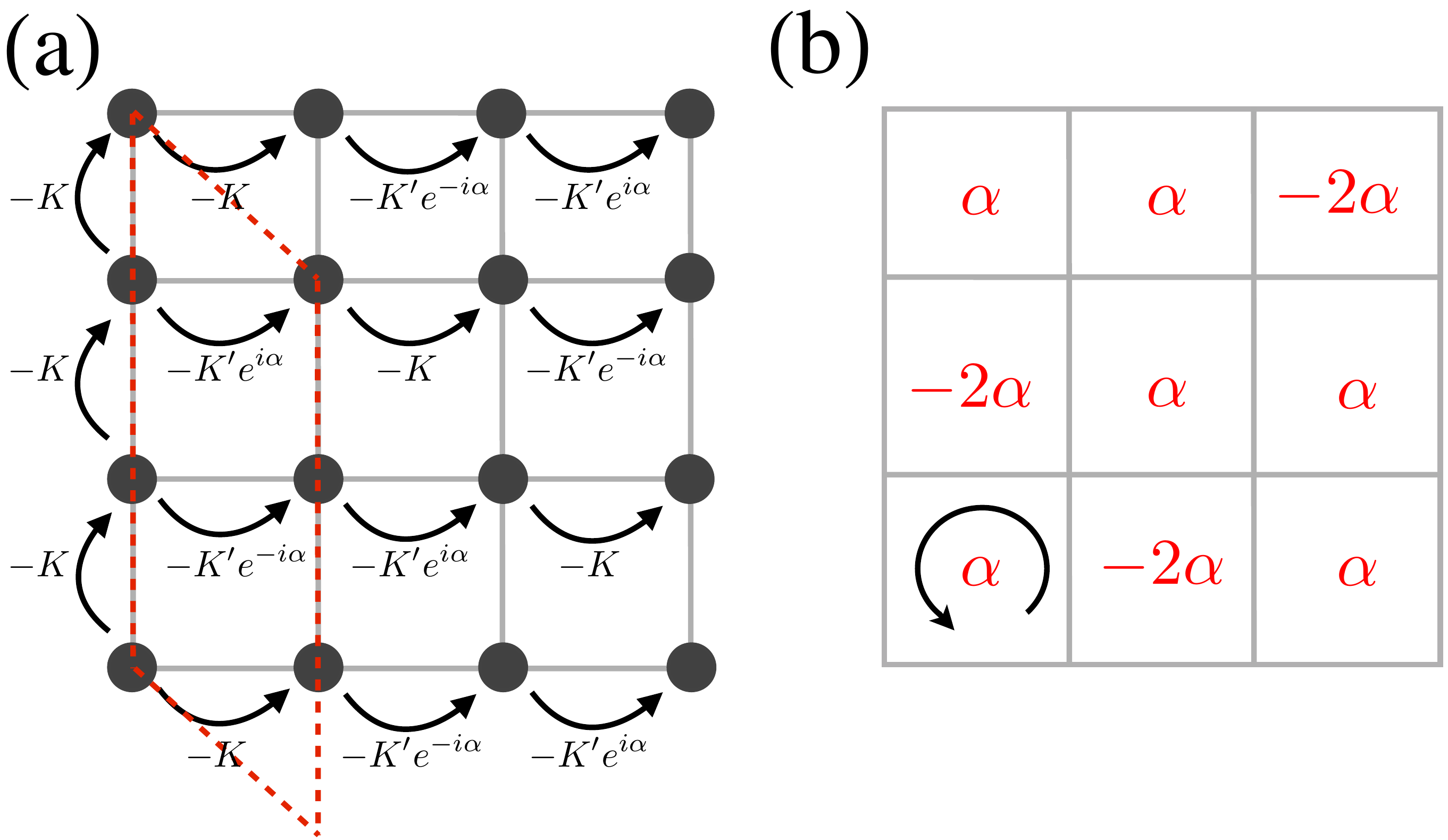}
\caption{(Color online) (a) Illustration of hopping matrix elements for the tight binding lattice model in the gauge described in the text. The area enclosed by the red dashed line is a unit cell containing three lattice sites. (b) Flux per plaquette for lattice shown in (a). When $\alpha=2\pi/3$, the flux through each plaquette is $1/3$ of an elementary flux quantum.}\label{fig:tightbinding}
\end{figure}

For dimensionless parameter $r=1$ this is simply the nearest neighbour
tight-binding model on a square lattice with no net flux.  (The phase $\theta$ can be removed by a gauge transformation.) The
ground state is a simple condensate without any flow. Reducing the control
parameter to $r=0$ smoothly interpolates to the Harper-Hofstadter
lattice model with flux $1/3$ per plaquette for which the ground state is a vortex lattice. Crucially, trajectories
can be found for $r$ in the range $1\geq r > 0$, such that there is always a
{\it unique} many-body ground state, and the system is adiabatically
transferred into a stable vortex lattice phase at $r=0$ with high
fidelity.  At the final point, $r=0$, the many-body state is
(six-fold) degenerate, corresponding to the different
translations/rotations of the vortex lattice. However energy barriers
of order $U$ per particle from interactions exist between these
states, preventing the formation of domains. Depending on the
trajectory, the system can be prepared in different translations of
the vortex lattice. For example, if we take $\theta=0$, the hopping
matrix elements take the form
\begin{eqnarray}
K_{n,m}=\left\{ \begin{array}{c} K \;\;\; m+n=0,3,\ldots\\ K'e^{-i \alpha} \;\; m+n=1,4,\ldots \\ K'e^{i \alpha} \;\;\; m+n=2,5,\ldots \end{array} \right. .
\end{eqnarray}
with $K'/K=\sqrt{1+3 r (1-r)}$, $\alpha=\arg[r+(1-r) e^{i
  2\pi/3}]$.
  
To demonstrate adiabaticity, in
Fig.~\ref{fig:currentstb}~(b) we show the sum of the magnitude of the
currents $J_{ij}$ between lattice sites $i$ and $j$ per bond
\begin{equation}
J=\frac{1}{N_{\rm bonds}}\sum_{\langle i,j \rangle} |J_{ij}|
\end{equation}
as a function of $1-r$.  (As above, we assume that interactions
are sufficiently weak that the atoms only occupy the lowest energy
single particle state, which is non-degenerate for $r\neq 0$.)  In the
lattice with uniform flux (i.e. $r=0$), these vortex lattice states
have $J_{ij}=\pm (3 K/2) N_{\rm site}$ along any bond with non-zero
current (here $N_{\rm site}$ denotes the average number of particles
per lattice site). We normalized $J$ in Fig. \ref{fig:currentstb} (b)
by its value at $r=0$, $J_0 \equiv N_{\rm site} K/2$. In the more
general case, while single-particle states are always non-degenerate,
different choices of the relative phase $\theta$ will load into one of
the three ground-states of the model with uniform flux shown in
Fig. \ref{fig:currentstb} (a)~\cite{footnote2}. The loading process is adiabatic as
long as the trajectory in $(r, \theta)$-space avoids crossing the
lines along $\theta = 2 \pi/6, \pi, -2\pi/6$ with $0<r<1/2$ where the
lowest energy single particle states changes discontinuously in
$\mathbf{k}$-space. These ground-states, shown in
Fig. \ref{fig:currentstb} (a), respect the (reduced) translation
symmetry of the unit cell in Fig. \ref{fig:tightbinding} (a) and are
related by translations by one lattice site.

\section{Conclusion}

We have described two protocols by which artificial gauge potentials
can be used to load cold bosonic atoms adiabatically into a vortex
lattice. In essence our strategy is to find ways by which the single
particle bandstructure interpolates between that of a free particle
(or simple, non-topological, bandstructure) and that of a particle in
an effective magnetic field. A BEC formed in the minimum of this band
can then be adiabatically transformed into a dense vortex lattice. As
we have emphasized, additional care is required when the final vortex
lattice breaks a symmetry of the system (e.g. spin-rotation, or
translation).  Then, to prevent domain formation and ensure adiabatic
loading, a route must be found which transfers the BEC directly into
one of these symmetry-broken phases. We have shown how this can be
done both for the optical flux lattice (by lifting the
spin-degeneracy) and for the tight-binding model (by using a route
which breaks the translational symmetry).

If loaded successfully into the desired ground-state vortex lattice
configurations, time-of-flight expansion after rapidly turning off the
optical lattice will reveal a pure condensate in a single momentum
state. In the first example, this will result in peaks at momenta
$(2i \mathbf{k}_1+j \mathbf{k}_3) \hbar t/M$ (with
$i,j=0,\pm1,\pm2, \ldots$) in the total density
$n(\mathbf{r})=n_{\uparrow}(\mathbf{r})+n_{\downarrow}(\mathbf{r})$
after expansion time $t$. In general, time-of-flight images of cold
atoms in artificial gauge potentials will be gauge
dependent~\cite{Lin:2009qf,Moller:2010uq,Kolovsky:2011zr}. Observing a
condensate in a single momentum state (rather than a linear combination of
the minima) indicates that the condensate has the same translation
symmetries as the implemented gauge. For example for the protocol for
loading into the tight-binding lattice Eq. \eqref{eq:matrixelments},
finding a condensate in one of the three degenerate minima of the
dispersion means that one of the three vortex lattice shown in
Fig. \ref{fig:currentstb} (a) was realized. The other three degenerate
mean-field ground-states are rotated by 90$^{\circ}$ and have a
different unit cell than the one shown in Fig. \ref{fig:tightbinding}
(a), and therefore are superpositions of the single-particle
states at the minima of the dispersion in this particular gauge. Other
detection techniques would naturally rely on detecting the density
wave associated with the vortex lattice, which could be detected by
in-situ probes such as as light scattering or single site resolution
imaging~\cite{Javanainen:1995bh,Corcovilos:2010qf,Bakr:2009dq,Sherson:2010cr}. 

Finally, we note that our protocols will also help to reach interesting 
regimes of strong correlations for bosons at high magnetic flux density. 
By first loading adiabatically into a dense vortex lattice at weak 
interaction strength and subsequently ramping to strong interactions, it 
may be possible to observe novel strongly correlated phases, e.g. 
fractional quantum Hall states of bosons in quasi-2D 
systems\cite{advances}, which typically require low entropies.

\acknowledgments{We are grateful to Monika Aidelsburger and Immanuel
  Bloch for helpful discussions and for sharing their unpublished
  ideas, and to Jean Dalibard for helpful comments. NRC thanks the Max Planck Institute of Quantum Optics for hospitality and the A. von Humboldt foundation for support. This work was supported by EPSRC EP/I010580/1.}


\begin{thebibliography}{10}
\bibitem{footnote1}
For example, for a shallow lattice $V_0/E_{\rm L}\ll 1$, the wavefunctions are
  $|k_+\rangle \simeq |\uparrow\rangle - (V_0/E_{\rm L})\, (c_1+ic_2)
  |\downarrow$ and $|k_- \rangle = \hat{T}_2 |k_+\rangle \simeq
  |\downarrow\rangle - (V_0/E_{\rm L}) \, (c_1 - i c_2) |\uparrow\rangle$,
  where $c_i \equiv \cos \mathbf{k}_i\cdot\mathbf{r}$.
\bibitem{footnote2}
The other three configurations would be formed by changing the photon assisted tunneling $K_{n,m}^0$, such that its phase is constant along lines of 
constant $m-n$.
\end{thebibliography}
\end{document}